\documentclass[12pt]{article}
\usepackage{makeidx}
\usepackage{amsmath}
\usepackage{amsfonts}
\usepackage{amssymb}
\usepackage{amsthm}
\usepackage{algorithm}
\usepackage{algpseudocode}
\usepackage{amsmath}
\usepackage{xcolor}
\usepackage{listings}
\usepackage{alltt}
\usepackage{tikz}
\usepackage{graphicx}
\usepackage{pgfplots}

\pgfplotsset{compat=1.16}

\def\1{{\bf 1}}

\begin{document}

\title{A conjecture on a tight norm inequality in the finite-dimensional $%
l_p $}
\author{A. S. Holevo, A. V. Utkin \\
Steklov Mathematical Institute, RAS, Moscow, Russia}
\date{}
\maketitle

\begin{abstract} We suggest a tight inequality for norms in $d$-dimensional space
$l_p $ which has simple formulation but appears hard to prove. We give a proof for $d=3$ and
provide a detailed numerical check for $d\leq 200$ confirming the conjecture. An unusual feature is the ``phase transition'' in
the constant of the inequality, depending on the dimension  $d$ and the parameter $p$. We conclude
with a brief survey of solutions for kin problems which anyhow concern
minimization of the output entropy of certain quantum channel and rely upon the symmetry
properties of the problem.

\end{abstract}

\bigskip

\section{Formulation of the problem}

Quantum information theory suggests a variety of optimization problems most
of which are hard to solve analytically. For problems such as computation of
the quantum channel capacity or accessible information the mathematical
difficulty is finding a global maximum of a convex function.
The problem considered in the present note
arose in connection with the computation of accessible information for the
ensemble of \textquotedblleft quantum pyramid\textquotedblright\ (see \cite%
{eng}) in \cite{ahut}. However it can be naturally formulated as
optimization problem in $d$-dimensional Banach space $l_{p}$ without any
reference to quantum information science.

Let $d\geq 3$ (the case $d=2$ is trivial) and consider the $(d-1)-$%
dimensional hyperplane
\begin{equation*}
L=\left\{ \mathbf{x}=(x_1,\dots ,x_d)\in\mathbb{R}^{d}: x_1+\dots +x_d=0 \right\} .
\end{equation*}

\textit{We conjecture the following tight inequalities: For }$\alpha \geq 1$%
\textit{\ }%
\begin{equation}
\left\Vert \mathbf{x}\right\Vert _{2\alpha }\leq M\left( d,\alpha \right)
^{1/2\alpha }\left\Vert \mathbf{x}\right\Vert _{2},\quad \mathbf{x}\in L;
\label{geq1}
\end{equation}%
\textit{For }$0<\alpha <1$\textit{\ }%
\begin{equation}
\left\Vert \mathbf{x}\right\Vert _{2\alpha }\geq M\left( d,\alpha \right)
^{1/2\alpha }\left\Vert \mathbf{x}\right\Vert _{2},\quad \mathbf{x}\in L,
\label{less1}
\end{equation}%
\textit{where the constant } $M\left( d,\alpha \right) $ \textit{is exact
and is defined as follows: }
\begin{equation}
M\left( d,\alpha \right) =\left\{
\begin{array}{cc}
2^{1-\alpha }, & d\leq d(\alpha ); \\
d^{-\alpha }\left[ \left( d-1\right) ^{\alpha }+\left( d-1\right) ^{1-\alpha
}\right] , & d>d(\alpha )%
\end{array}%
\right.  \label{mda}
\end{equation}%
\textit{when } $\alpha >1/2,$\textit{\ and }$M\left( d,\alpha \right)
=2^{1-\alpha }$ \textit{when} $\alpha \leq 1/2$ \textit{(in particular, $%
\left\Vert \mathbf{x}\right\Vert _{1}\geq \sqrt{2}\left\Vert \mathbf{x}%
\right\Vert _{2},\,\mathbf{x}\in L$). Here the value }$d(\alpha )$\textit{\
is the largest root of the equation}%
\begin{equation}
2^{1-\alpha }=d^{-\alpha }\left[ \left( d-1\right) ^{\alpha }+\left(
d-1\right) ^{1-\alpha }\right] .  \label{eq0}
\end{equation}

In the case $d\leq d(\alpha )$ the equality in the conjectured inequalities (%
\ref{geq1}), (\ref{less1}) is attained for $x_{1}=-x_{2}=1/\sqrt{2}%
,\,x_{j}=0 $, $j\geq 3;$ in the case $d>d(\alpha )$\ -- for $x_{1}=\sqrt{%
\frac{d-1}{d}},\,x_{j}=-\sqrt{\frac{1}{\left( d-1\right) d}},\,j\geq 2,$
(and for all permutations and total change of sign of such $x_{j}).$
Correspondingly, we speak of the maximizers of the \textit{first} and \textit{second} types.
\bigskip

For $\alpha >1/2$ the equation (\ref{eq0}) has exactly two roots: $d=2$ and $%
d=d(\alpha )>2.$ The function $d(\alpha )$ is monotonically decreasing from $%
+\infty $ for $\alpha =1/2$ to $d(1\mp 0)=6.47...$, which is the solution of
the equation
\begin{equation}
\log (d/2)=\frac{d-2}{d}\log (d-1),  \label{eqd}
\end{equation}%
and then to $d(\infty )=2.$ The equation (\ref{eqd}) is obtained by taking
the logarithmic derivative of (\ref{eq0}) with respect to $\alpha $ at $%
\alpha =1.$ Another important point is $\alpha =2$ for which $d(2)=3.$ For
all $\alpha >2$ it holds $d(\alpha )<3,$ and since the dimension $d$ takes
only integer values $3,4,\dots ,$ the quantity $M\left( d,\alpha \right) $
is then always given by the second option in (\ref{mda}).

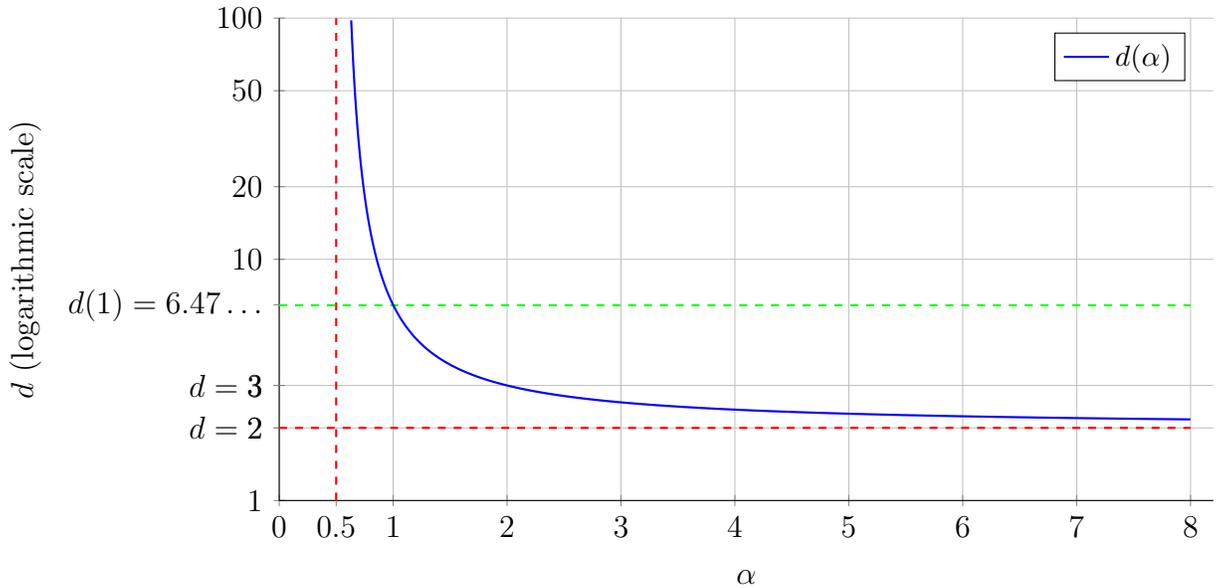
\begin{figure}[h!]
\label{Fig_d} \centering
\begin{tikzpicture}
\begin{axis}[
    width=14cm,
    height=8cm,
    xlabel={$\alpha$},
    ylabel={$d$ (logarithmic scale)},
    xmin=0, xmax=8.2,
    ymin=1, ymax=100,
    ymode=log,
    log basis y={10},
    xtick={0,1,2,3,4,5,6,7,8,9},
    xticklabels={0,1,2,3,4,5,6,7,8,9},
    extra x ticks={0.5},
    extra x tick labels={$0.5$},
    extra x tick style={grid=none},
    ytick={1,2,3,10,20,50,100},
    yticklabels={1,2,3,10,20,50,100},
    extra y ticks={2,3,6.47},
    extra y tick labels={$d=2$,$d=3$,$d(1)=6.47\ldots$},
    extra y tick style={grid=none},
    grid=both,
    grid style={line width=.1pt, draw=gray!30},
    major grid style={line width=.2pt, draw=gray!50},
    minor grid style={line width=.1pt, draw=gray!20},
    legend pos=north east,
    legend style={font=\small},
    axis lines=left,
    axis line style={-},
]

\draw[red, dashed, thick] (axis cs:0.5,1) -- (axis cs:0.5,100);

\draw[red, dashed, thick] (axis cs:0,2) -- (axis cs:8,2);

\draw[green, dashed, thick] (axis cs:0,6.45) -- (axis cs:8,6.45);

\addplot[blue, thick, mark=none] table[x=alpha, y=dalpha, col sep=space] {d_alpha_latex.txt};
\addlegendentry{$d(\alpha)$}

\end{axis}
\end{tikzpicture}
\caption{The largest root $d(\protect\alpha)$ of equation $d^{-\protect\alpha%
}\big((d-1)^{\protect\alpha}+(d-1)^{1-\protect\alpha}\big)=2^{1-\protect%
\alpha}$.}
\end{figure}

The plot of the function $\alpha \rightarrow d(\alpha ),\,\alpha >0,$ is
given in Fig. \ref{Fig_d}.

The proof of the conjectured inequalities reduces to the following problem:

\textit{For $d\geq 3$ and $\alpha \geq 1$ show that
\begin{equation}
M(d,\alpha )=\max_{\mathbf{x}}\sum_{j=1}^{d}\left\vert x_{j}\right\vert
^{2\alpha }  \label{ineqw}
\end{equation}%
under the constraints}
\begin{equation}
\sum_{j=1}^{d}\left\vert x_{j}\right\vert ^{2}=1,\quad \sum_{j=1}^{d}x_{j}=0,
\end{equation}%
\textit{with the maximizers described above.}

\textit{For } $\alpha <1$ \textit{the maximum in (\ref{ineqw}) is replaced
by the minimum. }


For $\mathbf{x\in }\mathbb{R}^{d}$ with $\left\Vert \mathbf{x}\right\Vert
_{2}=1$ introduce the probability distribution $P_{\mathbf{x}}=\left(
\left\vert x_{1}\right\vert ^{2},\dots ,\left\vert x_{d}\right\vert
^{2}\right) $ . Then for $\alpha\neq 1$ the problem can be reformulated in
terms of $\alpha $-R\'{e}nyi entropy $H_{\alpha }(P_{\mathbf{x}})=\left(
1-\alpha \right) ^{-1}\log \sum_{j=1}^{d}\left\vert x_{j}\right\vert
^{2\alpha }:$%
\begin{equation*}
\min_{\mathbf{x}\in L}H_{\alpha }(P_{\mathbf{x}})=\left\{
\begin{array}{cc}
\log 2, & d\leq d(\alpha ); \\
\left( 1-\alpha \right) ^{-1}\log d^{-\alpha }\left[ \left( d-1\right)
^{\alpha }+\left( d-1\right) ^{1-\alpha }\right] , & d>d(\alpha )%
\end{array}%
\right.
\end{equation*}%
The case $\alpha \rightarrow 1$ corresponds to minimization of the Shannon
entropy $H(P_{\mathbf{x}})$ considered in \cite{ahut} in connection with the
problem of accessible information for ensemble of ``quantum pyramid'' \cite%
{eng}. Going to the limit and taking into account that $6<d(1\pm 0)<7$
amounts to
\begin{equation*}
\min_{\mathbf{x}\in L}H(P_{\mathbf{x}})=\left\{
\begin{array}{cc}
\log 2, & d\leq 6; \\
\log d-\frac{d-2}{d}\log (d-1), & d\geq 7.%
\end{array}%
\right.
\end{equation*}

\section{Results for $d=3$}

Let us first focus on the case $\alpha >1$. Then the value $\alpha =2$ is of
special importance since $3=d(2).$ The hypothesis is that for $1<\alpha <2$
the maximum $M\left( 3,\alpha \right) =2^{1-\alpha }$ is attained on
(permutations of) $\left( 1/\sqrt{2},-1/\sqrt{2},0\right) $ and for $\alpha
>2,$ the maximum $M\left( 3,\alpha \right) =3^{-\alpha }\left( 2^{\alpha
}+2^{1-\alpha }\right) $ is attained on (permutations and total change of
sign of) $\left( \sqrt{2/3},-1/\sqrt{6},-1/\sqrt{6}\right) .$ The transition
for $\alpha =2$ between the two regimes is rather remarkable, namely:

For all $\mathbf{x}=(x_{1},x_{2},x_{3})$ satisfying%
\begin{equation}
x_{1}^{2}+x_{2}^{2}+x_{3}^{2}=1,\quad x_{1}+x_{2}+x_{3}=0  \label{cond}
\end{equation}%
it holds%
\begin{equation*}
x_{1}^{4}+x_{2}^{4}+x_{3}^{4}\equiv \frac{1}{2}=M\left( 3,2\right) .
\end{equation*}%
\textbf{Lemma.} \textit{For $\mathbf{x}$ satisfying (\ref{cond}) there is an
angle $\varphi $ such that%
\begin{equation}
x_{j}=\sqrt{2/3}\cos \left( \varphi +\frac{2\pi (j-1)}{3}\right) ,\quad
j=1,2,3.  \label{trineq}
\end{equation}%
}

\bigskip Note: $\varphi =\frac{\pi }{6}$ corresponds to the maximizer $%
\left(1/\sqrt{2},-1/\sqrt{2},0\right) ,\varphi =0$ corresponds to $\left(
\sqrt{2/3},-1/\sqrt{6},-1/\sqrt{6}\right) .$

\textit{Proof}. The conditions (\ref{cond}) mean that $\mathbf{x}$ is a unit
vector in the plane $L=\left\{ \mathbf{x}: x_{1}+x_{2}+x_{3}=0\right\} .$
Denote by $\mathbf{e}_{j}$ the coordinate orts, $\mathbf{\hat{e}}_{j}$ their
projections onto the plane $L,$ then $\mathbf{\hat{e}}_{1} =\left(
2/3,-1/3,-1/3\right) $ etc. with $\left\Vert \mathbf{\hat{e}}_{j}\right\Vert
=\sqrt{2/3}$ and
\begin{equation*}
x_{j}=\mathbf{e}_{j}\mathbf{x}^{\top }=\mathbf{\hat{e}}_{j}\mathbf{x}^{\top
}=\sqrt{2/3}\cos \varphi _{j},\quad j=1,2,3,
\end{equation*}%
where $\varphi _{j}$ is the angle between the vectors $\mathbf{x}$ and $%
\mathbf{\hat{e}}_{j}$ in the plane $L.$ The angle between any two different
vectors $\mathbf{\hat{e}}_{j}$ is $2\pi /3$ because $\mathbf{\hat{e}}_{j}%
\mathbf{\hat{e}}_{k}^{\top }=-\frac{1}{2}\left\Vert \mathbf{\hat{e}}%
_{j}\right\Vert \left\Vert \mathbf{\hat{e}}_{k}\right\Vert $. Therefore $%
\varphi _{j}=\varphi +\frac{2\pi (j-1)}{3},\quad j=1,2,3,$ thus we come to (%
\ref{trineq}). Then (\ref{cond}) become%
\begin{eqnarray}
\sum_{j=0}^{2}\cos \left( \varphi +\frac{2\pi j}{3}\right) &=&0
\label{cond0} \\
\frac{2}{3}\sum_{j=0}^{2}\cos ^{2}\left( \varphi +\frac{2\pi j}{3}\right)
&=&1  \label{cond1}
\end{eqnarray}%
and%
\begin{eqnarray*}
\sum_{j=1}^{3}x_{j}^{4} &=&\frac{4}{9}\sum_{j=0}^{2}\cos ^{4}\left( \varphi +%
\frac{2\pi j}{3}\right) \\
&=&\frac{1}{9}\sum_{j=0}^{2}\left[ 1+\cos \left( 2\varphi +\frac{4\pi j}{3}%
\right) \right] ^{2}=\frac{1}{2},
\end{eqnarray*}%
where we used (\ref{cond0}), (\ref{cond1}) with the doubled argument of
cosine (see also \eqref{aver} below).

\bigskip

\textbf{Theorem.} \textit{For $1<\alpha <2$ the maximum $M\left(
3,\alpha \right) =2^{1-\alpha }$ is attained on (permutations of) $\left( 1/%
\sqrt{2},-1/\sqrt{2},0\right) $ i.e. $\varphi =\frac{\pi }{6}.$ For $%
\alpha >2$ the maximum $M\left( 3,\alpha \right) =3^{-\alpha }\left(
2^{\alpha }+2^{1-\alpha }\right) $ is attained on (permutations of) $\left(
\sqrt{2/3},-1/\sqrt{6},-1/\sqrt{6}\right) $ i.e. $\varphi =0.$ }

\textit{For $0<\alpha <1$ the minimum $M\left( 3,\alpha \right) =2^{1-\alpha
}$ is attained on (permutations of) $\left( 1/\sqrt{2},-1/\sqrt{2},0\right) $%
.}

\textit{Proof.} For $0<\alpha <2$ we follow the method of \cite{sas} corresponding to the case $%
\alpha \rightarrow 1$ (minimum of the Shannon entropy).  With the aid of the Euler formula and the sum of geometric
progression, one has (in our case $m=3$):%
\begin{equation}  \label{aver}
\overline{\cos \left( \phi +L\frac{4\pi j}{m}\right) }\equiv \mathrm{Re~ e}%
^{i\phi}\frac{1}{m}\sum_{j=0}^{m-1}\exp{\left(iL\frac{4\pi j}{m}\right)}
=\left\{
\begin{array}{cc}
\cos \phi , & \frac{2L}{m}\text{ integer} \\
0 & \text{otherwise}%
\end{array}%
\right. ,
\end{equation}%
where bar denotes averaging over the values of $j$.

We first consider the case $1<\dot{\alpha}<2$ and the maximum of%
\begin{equation*}
M(\varphi )=\sum_{j=1}^{3}\left\vert x_{j}\right\vert ^{2\alpha }=\left(
\frac{2}{3}\right) ^{\alpha }\sum_{j=0}^{2}\left\vert \cos \left( \varphi +%
\frac{2\pi j}{3}\right) \right\vert ^{2\alpha }=3^{1-\alpha }\overline{\left[
1+\cos \left( 2\varphi +\frac{4\pi j}{3}\right) \right] ^{\alpha }}.
\end{equation*}%
For $\Vert \xi \Vert \leq 1$ we have
\begin{eqnarray}
\left( 1+\xi \right) ^{\alpha } &=&1+\alpha \xi +\sum_{n=2}^{\infty }\binom{%
\alpha }{n}\xi ^{n}  \label{exp} \\
&=&1+\alpha \xi +\sum_{n=1}^{\infty }c_{2n}\xi ^{2n}-\sum_{n=1}^{\infty
}c_{2n+1}\xi ^{2n+1},  \notag
\end{eqnarray}%
where $c_{2n}=\binom{\alpha }{2n},c_{2n+1}=-\binom{\alpha }{2n+1}$ are all
positive for $1<\alpha <2.$ Thus taking into account (\ref{cond0})%
\begin{eqnarray*}
M(\varphi ) &=&3^{1-\alpha }[1+\alpha \overline{\cos \left( 2\varphi +\frac{%
4\pi j}{3}\right) } \\
&+&\sum_{n=1}^{\infty }c_{2n}\overline{\cos \left( 2\varphi +\frac{4\pi j}{3}%
\right) ^{2n}}-\sum_{n=1}^{\infty }c_{2n+1}\overline{\cos \left( 2\varphi +%
\frac{4\pi j}{3}\right) ^{2n+1}}] \\
&=&3^{1-\alpha }\left[ 1+\sum_{n=1}^{\infty }c_{2n}\overline{\cos \left(
2\varphi +\frac{4\pi j}{3}\right) ^{2n}}-\sum_{n=1}^{\infty }c_{2n+1}%
\overline{\cos \left( 2\varphi +\frac{4\pi j}{3}\right) ^{2n+1}}\right] .
\end{eqnarray*}%
By using the formulas for the powers of the cosines (see \cite{rg},
n.1.320), 
\begin{equation}
\overline{\cos \left( 2\varphi +\frac{4\pi j}{3}\right) ^{2n}}=\frac{1}{%
2^{2n-1}}\left\{ \frac{1}{2}\binom{2n}{n}+\sum_{l=1}^{n}\binom{2n}{n-l}%
\overline{\cos \left[ 2l\left( 2\varphi +\frac{4\pi j}{3}\right) \right] }%
\right\}  \label{even}
\end{equation}%
\begin{equation}
\overline{\cos \left( 2\varphi +\frac{4\pi j}{3}\right) ^{2n+1}}=\frac{1}{%
2^{2n}}\left\{ \sum_{l=0}^{n}\binom{2n+1}{n-l}\overline{\cos \left[ \left(
2l+1\right) \left( 2\varphi +\frac{4\pi j}{3}\right) \right] }\right\}
\label{odd}
\end{equation}%
Then the averaging formula (\ref{aver}) implies
\begin{equation}
M(\varphi )=\sum_{l=0}^{\infty }\left[ \tilde{c}_{2l}\cos \left( 2l\right)
2\varphi -\tilde{c}_{2l+1}\cos \left( 2l+1\right) 2\varphi \right] ,
\label{exp1}
\end{equation}%
where all  coefficients with tilde are nonnegative. The maximal value of $%
M(\varphi )$ is attained if $\cos \left( 2l\right) 2\varphi =1,$ $\cos
\left( 2l+1\right) 2\varphi =-1,$ i.e $\varphi =\frac{\pi }{2}.$

On the contrary, in the case $0<\alpha < 1$ the coefficients with tilde are
nonpositive. Then the minimal value of $M(\varphi )$ is attained when $\cos
\left( 2l\right) 2\varphi =1,$ $\cos \left( 2l+1\right) 2\varphi =-1,$ i.e $%
\varphi =\frac{\pi }{2}.$

The function $M(\varphi )$ has period $\frac{\pi }{3},$ it is even with
respect to $\varphi =0$ and $\varphi =\frac{\pi }{6},$ therefore the value $%
\varphi =\frac{\pi }{2}$ corresponds to $\frac{\pi }{6}=\frac{\pi }{2}-\frac{%
\pi }{3}$.

In the case of integer $\alpha >2$ the power expansion (\ref{exp}) has
finite number of terms with all coefficients positive. Thus instead of (\ref%
{exp1}) we obtain
\begin{equation*}
M(\varphi )=\sum_{l=0}^{\alpha }\tilde{c}_{l}^{\prime }\cos 2\varphi l,
\end{equation*}%
with positive coefficients. This expression is maximized for $\varphi =0.$

Rather surprisingly, this approach does not have a simple extension to the case
of noninteger $\alpha >2$. Due to periodicity and evenness, it is sufficient to study
the behavior of $M(\varphi )$ on the segment $[0,\frac{\pi }{6}]$. In the Appendix,
we give the Fourier expansion of the function $M(\varphi )$.  Numerical study suggests that for $\alpha >2$
\begin{equation*}
-M^{\prime }(\varphi )/\sin 6\varphi \geq C(\alpha )>0,\quad \varphi \in
\left[ 0,\frac{\pi }{6}\right],
\end{equation*}%
hence $M(\varphi )$ is decreasing and the maximum is attained for $\varphi
=0.$  To prove the hypothesis for $\alpha>2$ it would be sufficient to
prove $M^{\prime }(\varphi )\leq 0$ on $(0,\frac{\pi }{6})$.


Instead we give a proof using completely
different approach. It is sufficient to show that for $\alpha>2$ the conditions (\ref{cond}) imply
\begin{equation*}
x_{1}^{2\alpha }+x_{2}^{2\alpha }+x_{3}^{2\alpha }\leq 3^{-\alpha }\left(
2^{\alpha }+2^{1-\alpha }\right) .
\end{equation*}

Without loss of generality, assume that $x_{2}\leq x_{1}\leq 0\leq
x_{3}=-(x_{1}+x_{2})$. By introducing the variable $x=\frac{x_{1}-x_{2}}{%
x_{1}+x_{2}}\in \lbrack 0,1]$ and using
\begin{equation*}
{1+\frac{x^{2}}{3}=\frac{4}{3}\frac{x_{1}^{2}+x_{2}^{2}+x_{1}x_{2}}{%
(x_{1}+x_{2})^{2}}=\frac{2}{3}\frac{x_{1}^{2}+x_{2}^{2}+x_{3}^{2}}{%
(x_{1}+x_{2})^{2}}=\frac{2/3}{(x_{1}+x_{2})^{2}},}
\end{equation*}%
{one obtains
\begin{equation*}
(1+x)^{2}+(1-x)^{2}+4^{\alpha }\leq 2^{2\alpha }\frac{1}{3^{\alpha }}\Bigl%
(2^{\alpha }+2^{1-\alpha }\Bigl)\Bigl(\frac{3}{2}(1+x^{2}/3)\Bigl)^{\alpha
}=(4^{\alpha }+2)\Bigl(1+\frac{x^{2}}{3}\Bigl)^{\alpha },
\end{equation*}%
or }%
\begin{equation}
g_{\alpha }(x)\leq 4^{\alpha }+2  \label{gineq}
\end{equation}%
{for all $x\in \lbrack 0,1]$, where}

\begin{equation}
g_{\alpha }(x)=\frac{(1+x)^{2\alpha }+(1-x)^{2\alpha }-2}{(1+\frac{x^{2}}{3}%
)^{\alpha }-1}
\end{equation}%
is a  nonnegative function defined on the segment $[0,1]$. The value $%
g_{\alpha }(1)$ is equal to $\dfrac{4^{\alpha }-2}{(4/3)^{\alpha }-1}$.
Since
\begin{equation}
(4^{\alpha }+2)((4/3)^{\alpha }-1)-(4^{\alpha }-2)=(4^{2\alpha }-4^{\alpha
+1})/3>0,\quad \alpha >2,
\end{equation}%
$g_{\alpha }(x)$ satisfies the inequality \eqref{gineq} on the right end of $%
[0,1]$. Let us prove that $g_{\alpha }(x)$ is monotonically increasing on $%
[0,1]$.

\textbf{Observation.} Let $a(x),\,b(x)$ be twice differentiable functions for $x\geq 0
$ such that  $a(0)=b(0)=0;\,b(x)\neq 0,x>0.$ Then%
\begin{equation*}
\left( \frac{a(x)}{b(x)}\right) ^{\prime }=\frac{b^{\prime
}(x)\int_{0}^{x}\left( \frac{a^{\prime }(y)}{b^{\prime }(y)}\right) ^{\prime
}b(y)dy}{b(x)^{2}}.
\end{equation*}%
Consequently, if $b(x)$ and $\frac{a^{\prime }(x)}{b^{\prime }(x)}$ are
increasing, then $\frac{a(x)}{b(x)}$ is increasing.

\textit{Proof:} Direct check integrating by parts.

Applying this observation to $g_{\alpha }(x)$ with $a(x)=(1+x)^{2\alpha }+(1-x)^{2\alpha
}-2,\,b(x)=(1+\frac{x^{2}}{3})^{\alpha }-1,$ it suffices to prove that the
function  $\frac{[(1+x)^{2\alpha }+(1-x)^{2\alpha }]^{\prime }}{%
[(1+x^{2}/3)^{\alpha }]^{\prime }}$ is increasing, and then, by using the
same observation, it is sufficient to prove that
\begin{equation*}
u(x):=\frac{[(1+x)^{2\alpha }+(1-x)^{2\alpha }]^{\prime \prime }}{%
[(1+x^{2}/3)^{\alpha }]^{\prime \prime }}=3(2\alpha -1)\frac{(1+x)^{2(\alpha
-1)}+(1-x)^{2(\alpha -1)}}{(1+x^{2}/3)^{(\alpha -2)}(1+(2\alpha -1)x^{2}/3)}
\end{equation*}
is increasing.

We have $u(x)=3(2\alpha-1)u_1(x)u_2(x)$, where
\begin{enumerate}
\item $u_1(x)=\frac{(1+x)^{2(\alpha-1)}+(1-x)^{2(\alpha-1)}}{((1+x)^2+(1-x)^2)^{\alpha-1}}$;
\item $u_2(x)=\frac{((1+x)^2+(1-x)^2)^{\alpha-1}}{(1+x^2/3)^{(\alpha-2)}(1+(\alpha-1)x^2/3)}$.
\end{enumerate}
By making a monotonically increasing smooth substitution $t=\Bigl(\frac{1+x}{1-x}\Bigr)^2,\ t\in[1,+\infty)$,
the first factor $u_1(x)$ transforms into $\frac{1+t^{\alpha-1}}{(1+t)^{\alpha-1}}$.
Since the derivative of this expression with respect to $t$ equals $ (\alpha-1)\frac{t^{\alpha-2}-1}{(1+t)^{\alpha}}$
and is nonnegative for $\alpha\geq 2$, the function $u_1(x)$ is increasing in $x\in[0,1]$ (as a composition of increasing functions).

The logarithmic derivative of $u_2(x)$ is
\begin{equation*}
(\ln u_2(x))'=2x\Bigl(\frac{\alpha-1}{1+x^2}-\frac{\alpha-2}{3+x^2}-\frac{2\alpha-1}{3+(2\alpha-1)x^2}\Bigr)=\frac{8x^3(\alpha-1)(\alpha-2)}{(1+x^2)(3+x^2)(3+(2\alpha-1)x^2)},
\end{equation*}
which is also nonnegative. Hence, $u_2(x)$ is non-decreasing as well.
Hence $u(x)=3(2\alpha-1)u_1(x)u_2(x)$ is increasing,
and we have shown that $g'_{\alpha}(x)\geq 0$ on $[0,1]$,  which completes the proof.

%
%



\section{Numerical verification}

\paragraph{Theoretical background.}

We consider both cases $\alpha>1$ and $0<\alpha<1$.

\label{ss_th} The Lagrange method is useful to make the computation faster,
since the constraints (\ref{cond}) yield $O(d^2)$ one-dimensional
optimization problems in certain dimension $d$.

The necessary condition for the point $\mathbf{x}$ to be a critical point of
the Lagrange function
\begin{equation*}
L(\mathbf{x},\lambda)=\sum\limits_{j=1}^d |x_j|^{2\alpha}-\lambda(\|\mathbf{x%
}\|_2^2-1)-\mu(1\mathbf{x}^T)
\end{equation*}
is
\begin{equation*}
\dfrac{\partial L}{\partial x_j}(\mathbf{x})=2\alpha
x_j|x_j|^{2(\alpha-1)}-2\lambda x_j-\mu=0,\quad \forall :\ 1\leq j\leq d.
\end{equation*}
Thus all the coordinates of $\mathbf{x}$ satisfy the equation
\begin{equation}  \label{eq1}
2\alpha x|x|^{2(\alpha-1)}-2\lambda x-\mu=0
\end{equation}
for some real $\lambda$ and $\mu$, which has at most $3$ solutions. Indeed,
the derivative of the left hand side of \eqref{eq1} multiplied by $(1-\alpha)
$
\begin{equation}
4\alpha(\alpha-1)^2|x|^{2(\alpha-1)}-2\lambda(\alpha-1)
\end{equation}
is either nonnegative everywhere or negative on an interval.

This means that, up to permutation and sign changes, maximizer has the form
\begin{equation}
\mathbf{x}=(\underbrace{s_0,...,s_0}_{k_0\text{ times}},\underbrace{%
s_1,...,s_1}_{k_1\text{ times}},\underbrace{s_2,...,s_2}_{k_2\text{ times}%
}),\quad s_0\leq 0\leq s_1\leq s_2.
\end{equation}

To construct a one-dimensional parametrization put $s_0=-(k_1s_1+k_2s_2)/k_0$
into the quadratic form $q(s_1,s_2)=k_0s_0^2(s_1,s_2)+k_1s_1^2+k_2s_2^2$ and
find a linear transformation
\begin{equation}
\begin{pmatrix}
s_1 \\
s_2%
\end{pmatrix}%
=%
\begin{pmatrix}
u_1 &  & u_2 \\
v_1 &  & v_2%
\end{pmatrix}%
\begin{pmatrix}
c_1 \\
c_2%
\end{pmatrix}%
\end{equation}
such that $q(s_1(c_1,c_2),s_2(c_1,c_2))=\lambda_1 c_1^2+\lambda_2 c_2^2$.
The parameters $u_1,u_2,v_1,v_2$ can be calculated as follows:

\begin{enumerate}
\item Find coefficients $A,B,C$ of the form $%
q(s_1,s_2)=As_1^2+2Bs_1s_2+Cs_2^2$.
\begin{eqnarray*}
A=\dfrac{k_1(k_1+k_0)}{k_0} \\
B=\dfrac{k_1 k_2}{k_0} \\
C=\dfrac{k_2(k_2+k_0)}{k_0};
\end{eqnarray*}

\item Compute $\lambda_{1,2}=\dfrac{k_0(k_1+k_2)+k_1^2+k_2^2\pm\sqrt{D}}{2k_0%
}$, where $D=(k_0(k_1+k_2)+k_1^2+k_2^2)^2-4k_0k_1k_2d$;

\item Express
\begin{eqnarray*}
u_1=\dfrac{B}{\sqrt{B^2+(A-\lambda_1)^2}}, \\
v_1=\dfrac{\lambda_1-A}{\sqrt{B^2+(A-\lambda_1)^2}}
\end{eqnarray*}
and $u_2=-v_1,\ v_2=u_1$.

\item Then we find
\begin{equation*}
s_1=u_1c_1+u_2c_2,\quad s_2=v_1c_1+v_2c_2.
\end{equation*}
\end{enumerate}

Therefore, the set of pairs $(s_1,s_2)$ with condition $q(s_1,s_2)=1$ is
parametrized by $t\in[0,2\pi)$ according to the formula
\begin{equation}
\begin{pmatrix}
s_1(t) \\
s_2(t)%
\end{pmatrix}%
=%
\begin{pmatrix}
u_1 &  & u_2 \\
v_1 &  & v_2%
\end{pmatrix}%
\begin{pmatrix}
\frac{\cos(t)}{\sqrt{\lambda_1}} \\
\frac{\sin(t)}{\sqrt{\lambda_2}}%
\end{pmatrix}%
\end{equation}

Thus, when the maximizer of $F_\alpha(\mathbf{x})=\sum\limits_{j=1}^d
|x_j|^{2\alpha}$ for $\alpha>1$ (or the minimizer for $\alpha\in(0,1)$) has
non-zero coordinates, these coordinates must be equal to one of the
following real numbers -- $s_0$, $s_1$ or $s_2$ with multiplicities $k_0$, $%
k_1$ and $k_2=d-k_0-k_1$ -- the maximum $M(d,\alpha)$ can be found
numerically by considering the functions of one variable $f_{d,%
\alpha;k_1,k_2}(t)=k_0\Bigl(\frac{k_1s_1(t)+k_2s_2(t)}{k_0}%
\Bigl)^{2\alpha}+k_1s_1(t)+k_2s_2(t)$.


\paragraph{Numerical verification algorithm.}

We consider the maximization problem for the function
\begin{equation*}
F_{\alpha}(\mathbf{x}) = \sum_{j=1}^{d} |x_j|^{2\alpha}, \quad \alpha > 1,
\end{equation*}
under the constraints
\begin{equation*}
\|\mathbf{x}\|_2^2 = \sum_{j=1}^{d} x_j^2 = 1, \qquad \sum_{j=1}^{d} x_j = 0.
\end{equation*}

\begin{algorithm}[H]
\caption{Numerical verification of the maximization hypothesis for $F_\alpha$}
\begin{algorithmic}[1]

\State \textbf{Input:} Parameter $\alpha > 1$, dimension range $[d_{\min}, d_{\max}]$, tolerance $\epsilon = 10^{-8}$
\State \textbf{Output:} Verification result for each dimension $d$

\Procedure{VerifyHypothesis}{$\alpha$, $d_{\min}$, $d_{\max}$}
    \For{$d = d_{\min}$ \textbf{to} $d_{\max}$}

        \State \Comment{Compute theoretical bounds}
        \State $M_1(\alpha,d) \gets d^{-\alpha}\big((d-1)^\alpha + (d-1)^{1-\alpha}\big)$
        \State $M_2(\alpha,d) \gets 2^{1-\alpha}$
        \State $M_{\text{num}}(\alpha,d) \gets 0$

        \ForAll{ordered triples $(k_0,k_1,k_2)$ with $k_0+k_1+k_2 = d$, $k_0 \ge 1$, $k_1 \le k_2$}

            \State \Comment{Compute parametrization $s_0(t), s_1(t), s_2(t)$ as in Paragraph \ref{ss_th}}
            \State $(s_0(t), s_1(t), s_2(t)) \gets \text{Parametrization}(k_0,k_1,k_2)$

            \State \Comment{Maximize one-dimensional function}
            \State $f(t) \gets k_0|s_0(t)|^{2\alpha} + k_1|s_1(t)|^{2\alpha} + k_2|s_2(t)|^{2\alpha}$
            \State $M_{\text{cand}} \gets \max\limits_{t\in[0,2\pi)} f(t)$

            \If{$M_{\text{cand}} > M_{\text{num}}(\alpha,d)$}
                \State $M_{\text{num}}(\alpha,d) \gets M_{\text{cand}}$
            \EndIf
        \EndFor

        \State \Comment{Test hypotheses}
        \State $\Delta_1 \gets |M_{\text{num}}(\alpha,d) - M_1(\alpha,d)|$
        \State $\Delta_2 \gets |M_{\text{num}}(\alpha,d) - M_2(\alpha,d)|$
        \State $\text{valid}_1 \gets (\Delta_1 \leq \epsilon)$
        \State $\text{valid}_2 \gets (\Delta_2 \leq \epsilon)$

        \State \Comment{Accept hypothesis if either bound matches}
        \State $\text{confirmed} \gets \text{valid}_1 \ \textbf{or} \ \text{valid}_2$
        \State Output result for dimension $d$
    \EndFor
\EndProcedure

\end{algorithmic}
\end{algorithm}

For the numerical verification of hypothesis \eqref{less1} (with $%
\alpha\in(0,1)$), the algorithm description should involve finding the
minimum of the function $f(t)$ from the algorithm instead of its maximum.

\paragraph{Complexity.}

Instead of optimizing over $\mathbb{R}^d$, the algorithm reduces the problem
to $O(d^2)$ one-dimensional optimizations, making verification feasible for $%
d$ up to several hundred.

\paragraph{Expected outcome.}

The behavior of $M_{\text{num}}(d)$ depends on $\alpha$ as follows:

\begin{itemize}
\item For $\alpha>1$: there exists a critical dimension $d(\alpha)$ such
that
\begin{equation*}
M_{\text{num}}(d) \approx
\begin{cases}
2^{1-\alpha}, & d < d(\alpha) \\
d^{-\alpha}\big((d-1)^\alpha+(d-1)^{1-\alpha}\big), & d \geq d(\alpha)%
\end{cases}
\end{equation*}

\item For $0.5<\alpha<1$: the behavior is reversed, i.e.,
\begin{equation*}
M_{\text{num}}(d) \approx
\begin{cases}
d^{-\alpha}\big((d-1)^\alpha+(d-1)^{1-\alpha}\big), & d < d(\alpha) \\
2^{1-\alpha}, & d \geq d(\alpha)%
\end{cases}
\end{equation*}

\item For $0<\alpha\leq 0.5$: $M_{\text{num}}(d) \approx 2^{1-\alpha}$ for
all $d$.
\end{itemize}

The critical dimension satisfies \eqref{eqd}
\begin{equation*}
2^{1-\alpha} = d(\alpha)^{-\alpha}\big((d(\alpha)-1)^\alpha+(d(%
\alpha)-1)^{1-\alpha}\big).
\end{equation*}

\paragraph{Verification results.}

The algorithm confirms the hypothesis for all tested dimensions within
tolerance $\epsilon$, showing perfect agreement with the theoretical
predictions. Specifically, numerical verification was performed for $\alpha
= 0.05$, $0.2$, $0.45$, $0.5$, $0.55$, $0.7$, $0.95$, $1.01$, $1.1$, $1.5$,
and $2$ across dimensions $d = 3$ through $200$. In each case, the computed
maximum $M_{\text{num}}(d,\alpha)$ coincides (within the prescribed
tolerance) with the conjectured value. The correspondence holds for every
tested pair $(\alpha,d)$, confirming the validity of the structural
hypothesis over the investigated range.

\section{Discussion}
Here we give references to solutions of several problems akin to our conjecture. All cases one
way or another concern
minimization of the output entropy of certain quantum channel (normalized completely
positive map) and rely upon the symmetry properties of the problem.

In \cite{lieb1} Lieb gave a solution of the Wehrl conjecture which can be reformulated
as a conjecture about the minimal output entropy of the measurement (quantum-classical)
channel associated with the Glauber's coherent states with the underlying Heisenberg group,
and generalized the conjecture to SU(2) group.
The solution was based on the sharp versions of Young's and Hausdorff-Young inequalities in
the classical harmonic analysis. In \cite{lieb2}
Lieb and Solovej proved the Wehrl-type entropy conjecture
for symmetric SU(N) coherent states and suggested a similar conjecture for larger class
of Lie groups and their representation (for further progress in this direction see \cite{frank}
and the references therein). In \cite{lieb2} the authors used the ``universal cloning channel''
and established minimization for arbitrary concave function of the output distribution (the majorization).

Another relevant case is the solution of
the Gaussian optimizers conjecture for the classical capacity of bosonic
Gaussian channels by Giovannetti, Holevo and Garcia-Patron \cite{ghg}. The result for the minimal Wehrl entropy
problem can be obtained from a special limiting case of this \cite{ghm}. For one mode Gaussian measurement channels 
the aforementioned conjecture was settled in \cite{log1}, \cite{log2}.
In that case certain generalizations of the logarithmic Sobolev inequality
were used. In all these cases the symmetry group was a Lie group, while a generalization
to the case of Weyl system, associated with arbitrary (not necessarily continuous)
locally compact Abelian group paired with its dual, was elaborated by Zelenov \cite{zel}.

We surmise that the hypothesis of the present paper could be regarded as a discrete relative of
the aforementioned problems, within the context of the symmetric group and its standard representation.
It seems that maximizers in the problem \eqref{ineqw} can play the role of a discrete analogue of coherent state vectors.
However, what is unusual as compared to problems with continuous symmetry groups is the presence of two types of maximizers and a ``phase transition'' between them.
\bigskip

\noindent\textbf{Acknowledgment.} The authors are grateful to E.I. Zelenov for the reference to \cite{frank} and to other related works.

\section*{Appendix}

By using the expansions (\ref{even}), (\ref{odd}) with the interchanged
summation order and the averaging formula (\ref{aver}) one can obtain the
Fourier expansion
\begin{equation}
M(\varphi )=3^{1-\alpha }\left[ 1+C_{0}(\alpha )+\sum_{k=1}^{\infty }\cos
6k\varphi \sum_{n=0}^{\infty }\frac{\alpha (\alpha -1)\dots (\alpha -2n-3k+1)%
}{n!(n+3k)!}2^{-(2n+3k-1)}\right] ,  \label{mexp}
\end{equation}%
where $C_{0}(\alpha )=\sum_{n=1}^{\infty }\frac{\alpha (\alpha -1)\dots
(\alpha -2n+1)}{\left( n!\right) ^{2}}2^{-2n}.$

\textit{Proof}. Inserting (\ref{even}), (\ref{odd}) into the expansion (\ref%
{exp}), we obtain
\begin{equation*}
\begin{array}{c}
3^{\alpha -1}M(\varphi )=1+\sum_{n=1}^{\infty }\binom{\alpha }{2n}\frac{1}{%
2^{2n-1}}\left\{ \frac{1}{2}\binom{2n}{n}+\sum_{l=1}^{n}\binom{2n}{n-l}%
\overline{\cos \left[ 2l\left( 2\varphi +\frac{4\pi j}{3}\right) \right] }%
\right\} \\
+\sum_{n=1}^{\infty }\binom{\alpha }{2n+1}\frac{1}{2^{2n}}\sum_{l=1}^{n}%
\binom{2n+1}{n-l}\overline{\cos \left[ \left( 2l+1\right) \left( 2\varphi +%
\frac{4\pi j}{3}\right) \right] } \\
=1+C_{0}(\alpha )+\sum_{n=1}^{\infty }\sum_{l=1}^{n}\frac{\alpha (\alpha
-1)\dots (\alpha -2n+1)}{\left( n-l\right) !(n+l)!}2^{-(2n-1)}\overline{\cos %
\left[ 2l\left( 2\varphi +\frac{4\pi j}{3}\right) \right] } \\
+\sum_{n=1}^{\infty }\sum_{l=0}^{n}\frac{\alpha (\alpha -1)\dots (\alpha -2n)%
}{\left( n-l\right) !(n+l+1)!}2^{-2n}\overline{\cos \left[ \left(
2l+1\right) \left( 2\varphi +\frac{4\pi j}{3}\right) \right] } \\
=1+C_{0}(\alpha )+\sum_{l=1}^{\infty }\overline{\cos \left[ 2l\left(
2\varphi +\frac{4\pi j}{3}\right) \right] }\sum_{n=l}^{\infty }\frac{\alpha
(\alpha -1)\dots (\alpha -2n+1)}{\left( n-l\right) !(n+l)!}2^{-(2n-1)} \\
+\sum_{l=0}^{\infty }\overline{\cos \left[ \left( 2l+1\right) \left(
2\varphi +\frac{4\pi j}{3}\right) \right] }\sum_{n=l}^{\infty }\frac{\alpha
(\alpha -1)\dots (\alpha -2n)}{\left( n-l\right) !(n+l+1)!}2^{-2n}%
\end{array}%
\end{equation*}%
In the last sum the term with $l=0$ vanishes due to (\ref{aver}) and
introducing $n^{\prime }=n-l$ we obtain%
\begin{eqnarray*}
&&1+C_{0}(\alpha )+\sum_{l=1}^{\infty }\overline{\cos \left[ 2l\left(
2\varphi +\frac{4\pi j}{3}\right) \right] }\sum_{n^{\prime }=0}^{\infty }%
\frac{\alpha (\alpha -1)\dots (\alpha -2n^{\prime }-2l+1)}{\left( n^{\prime
}\right) !(n^{\prime }+2l)!}2^{-(2n^{\prime }+2l-1)} \\
&&+\sum_{l=1}^{\infty }\overline{\cos \left[ \left( 2l+1\right) \left(
2\varphi +\frac{4\pi j}{3}\right) \right] }\sum_{n^{\prime }=0}^{\infty }%
\frac{\alpha (\alpha -1)\dots (\alpha -2n^{\prime }-2l)}{\left( n^{\prime
}\right) !(n^{\prime }+2l+1)!}2^{-(2n+2l)} \\
&=&1+C_{0}(\alpha )+\sum_{L=1}^{\infty }\overline{\cos \left[ L\left(
2\varphi +\frac{4\pi j}{3}\right) \right] }\sum_{n^{\prime }=0}^{\infty }%
\frac{\alpha (\alpha -1)\dots (\alpha -2n^{\prime }-L+1)}{\left( n^{\prime
}\right) !(n^{\prime }+L)!}2^{-(2n^{\prime }+L-1)}
\end{eqnarray*}%
Due to (\ref{aver}), only terms corresponding to $L=3k,\,k=1,2,...$ survive,
hence (\ref{mexp}) follows.

\end{document}